\documentclass[prd, twocolumn, groupedaddress]{revtex4}

\usepackage{color}

\definecolor{myred}{rgb}{1,0,0}
\definecolor{mygreen}{rgb}{0,0.8,0.2}
\definecolor{myblue}{rgb}{0,0,1}

\usepackage{epsfig}
\usepackage{graphicx}
\usepackage{epstopdf}
\usepackage{amsmath}
\usepackage{amssymb}
\usepackage{amsfonts}

\def\slashchar#1{\setbox0=\hbox{$#1$}     		% set a box for #1
   \dimen0=\wd0                                 	% and get its size
   \setbox1=\hbox{/} \dimen1=\wd1               	% get size of /
   \ifdim\dimen0>\dimen1                        	% #1 is bigger
      \rlap{\hbox to \dimen0{\hfil/\hfil}}      	% so center / in box
      #1                                        	% and print #1
   \else                                        	% / is bigger
      \rlap{\hbox to \dimen1{\hfil$#1$\hfil}}   	% so center #1
      /                                         	% and print /
   \fi}

\begin{document}

\title{Pion condensation in a dense neutrino gas}

\author{Hiroaki Abuki} \email{abuki@th.physik.uni-frankfurt.de}
\affiliation{Institut f\"ur Theoretische Physik, Goethe-Universit\"at
  Frankfurt, Max-von-Laue-Stra\ss e~1, 60438 Frankfurt am Main, Germany}

\author{Tom\'a\v{s} Brauner}
\email{brauner@th.physik.uni-frankfurt.de}
\thanks{On leave from Department of Theoretical Physics,
Nuclear Physics Institute ASCR, 25068 \v Re\v z, Czech Republic.}
\affiliation{Institut f\"ur Theoretische Physik, Goethe-Universit\"at
  Frankfurt, Max-von-Laue-Stra\ss e~1, 60438 Frankfurt am Main, Germany}

\author{Harmen J. Warringa}
\email{warringa@th.physik.uni-frankfurt.de}
\affiliation{Institut f\"ur Theoretische Physik, Goethe-Universit\"at
  Frankfurt, Max-von-Laue-Stra\ss e~1, 60438 Frankfurt am Main, Germany}

\begin{abstract}
We argue that using an equilibrated gas of neutrinos it is
  possible to probe the phase diagram of QCD for finite isospin and
  small baryon chemical potentials.  We discuss this region of the
  phase diagram in detail and demonstrate that for large enough
  neutrino densities a Bose-Einstein condensate of positively charged
  pions arises. Moreover, we show that for nonzero neutrino density
  the degeneracy in the lifetimes and masses of the charged pions is
  lifted.
\end{abstract}

\date{August 26, 2009}

\maketitle

\section{Introduction}
Quantum chromodynamics (QCD) predicts that the properties of hadronic
matter hugely depend on the baryon and isospin density.  For example,
it is found that for small baryon chemical potential such that there
are no baryons around a Bose-Einstein condensate of charged pions
should arise if the absolute value of the isospin chemical potential
($\mu_I$) becomes larger than the vacuum pion mass $(m_\pi)$
\cite{Son01}.  This has been shown using the chiral effective
Lagrangian \cite{Son01, Kogut01, Loewe03}.  Furthermore the existence
of a pion condensate was also demonstrated in lattice QCD
\cite{Kogut02, Nishida04, Kogut04, deForcrand07, Detmold08}, with
random matrix theory \cite{Klein03}, using the Nambu--Jona-Lasinio
(NJL) model \cite{Toublan03, Barducci04, He05, He05b, Warringa05},
with the $O(4)$ linear sigma model \cite{Mao06, Andersen06} and with
string theory inspired holographic models of QCD \cite{Parnachev07,
  Erdmenger08, Rebhan08}.

The question we would like to address in this article is under which
conditions a pion condensate can be realized in a macroscopic system
of particles. In order for a macroscopic system to be stable, it
should be electrically neutral. Since the isospin chemical potential
is equal to the charge chemical potential ($\mu_I = \mu_Q$), requiring
neutrality puts restrictions to the value of the isospin chemical
potential.

At low baryon chemical potential ($\mu_B \lesssim m_p$, here $m_p$
denotes the mass of the proton), the electric neutrality constraint
forces $\mu_Q \approx 0$ so that pion condensation becomes impossible
\cite{Abuki08} (see also Refs.~\cite{Ebert06, Andersen07}). If however
a finite density of neutrinos which is in equilibrium with hadronic
matter is present, these considerations will be modified.  As we will
discuss in Sec.~II a nonzero density of neutrinos always increases
$\mu_Q$. If the baryon chemical potential is small and the chemical
potential of the neutrinos becomes large enough, it will be
energetically favored to convert some of the neutrinos into pions and
electrons. In this way as we will show in Sec.~II a pion condensate is
formed in a dense neutrino gas for low baryon chemical potential.
This has interesting consequences for the phase diagram of QCD at
finite electron and/or muon lepton number chemical potential.  In
Sec.~II and Sec.~III we will show that pion condensation takes place
in a large part of the phase diagram of QCD for $\vert \mu_B \vert
\lesssim m_p - m_\pi$.  Even before the onset of pion condensation,
the neutrino gas has interesting implications on the behavior of the
pions. Because of the nonzero isospin chemical potential, the
degeneracy in the masses of the charged pions will be lifted.
Moreover, the lifetime of the pions will change as we will see in
Sec.~IV.

Let us point out that similar condensation phenomena can also arise at
larger baryon chemical potential ($\mu_B \gtrsim m_p$).  In dense
electrically neutral nuclear matter, $-\mu_Q$ can become of the order
of the vacuum pion mass.  Hence condensation of negatively charged
pions in dense nuclear matter seemed a realistic possibility
\cite{Migdal72, Scalapino72, Sawyer72}.  However, due to interactions
the in-medium pion mass in nuclear matter is increased such that pion
condensation becomes improbable. On the other hand, the mass of the
negatively charged kaon is decreased due to an attractive interaction.
A negatively charged kaon condensate turns out to be a more likely
possibility in dense neutral nuclear matter \cite{Kaplan86} even at
high neutrino densities \cite{Pons00}.  In color superconducting
matter which is formed at even higher baryon densities, the masses of
the kaonic excitations are smaller than those of the pionic
excitations \cite{Son00}. It turns out that in electrically neutral
color superconducting matter a neutral kaon condensate can be formed,
creating the so-called CFL-$K^0$ phase \cite{Schafer00, Steiner02,
  Forbes05, Buballa05, Warringa06, Andersen08, Phat08}.

In nature a dense neutrino gas is created during a supernova
explosion. Part of the neutrinos produced in this explosion will be
trapped in the precursor of a neutron star, the so-called
proto-neutron star \cite{Prakash96}. Inside a proto-neutron star the
baryon density is very large, which as we stated above makes pion
condensation unlikely. However, one could speculate that in the crust
or in the atmosphere the neutrino density might be large and baryon
density small enough for pion condensation to occur. To fully
settle this issue one should also show that chemical and thermal
equilibrium can be reached.  In this paper, we will not give a
definite answer to the question whether or not a pion condensate is
realized somewhere in our universe. We will only unambiguously show
that a pion condensate is formed if the conditions are right, that is
in a dense neutrino gas at low baryon densities.

\section{Pion condensation}
At very large neutrino density and small baryon chemical potential it
is energetically favored to convert some of the neutrinos into pions
and leptons. In this section we will show that in such a case a pion
condensate will be formed. We will compute at which neutrino density
this will happen and discuss the phase diagram as a function of
electron and muon lepton number chemical potential.

In order to describe matter at finite density we express the chemical
potentials of the up quark ($u$), down quark ($d$), electron ($e$),
electron neutrino ($\nu_e$), muon ($\mu$) and muon neutrino
($\nu_\mu$) in terms of the chemical potentials of conserved
quantities. We will assume that chemical equilibrium is reached, hence
\begin{equation}
\begin{split}
\mu_u & = \tfrac{1}{3} \mu_B + \tfrac{2}{3} \mu_Q,
\\
\mu_d &= \tfrac{1}{3} \mu_B - \tfrac{1}{3} \mu_Q,
\\
\mu_e &= -\mu_Q + \mu_{L_e},
\\
\mu_{\nu_e} &= \mu_{L_e},
\\
\mu_{\mu} &= -\mu_Q + \mu_{L_\mu},
\\
\mu_{\nu_\mu} &= \mu_{L_\mu}.
\end{split}
\label{eq:chempot}
\end{equation}
Here $\mu_B, \mu_Q, \mu_{L_e}$ and $\mu_{L_\mu}$ denote respectively
the chemical potentials for baryon number, electric charge, electron
lepton number and muon lepton number. Strictly speaking, the muon and
electron lepton numbers are not separately conserved due to neutrino
oscillations. We will assume that the size of our system is much
smaller than the neutrino oscillation length.  Furthermore, we will
assume that the time it takes to bring the system in chemical
equilibrium is much shorter than the typical neutrino oscillation
time.  In this situation it is to first approximation correct to
introduce separate chemical potentials for electron and muon lepton
numbers.  On long time scales only total lepton number is
conserved and one can no longer talk about separate lepton number
chemical potentials. This situation is a special case of our results
in which $\mu_{L_e} = \mu_{L_\mu}$.

To make the system of particles electrically neutral, the charge
chemical potential has to be chosen in such a way that the charge
density $n_Q$ vanishes. This can be achieved by solving the following
equation
\begin{equation}
 n_Q = - \frac{\partial \Omega}{\partial \mu_Q} = 0,
\label{eq:neutrality}
\end{equation}
where $\Omega$ denotes the thermodynamic potential. It has a
hadronic and leptonic component. We will write
\begin{equation}
 \Omega = \sum_{i=e, \mu} \left( \Omega_i + \Omega_{\nu_i} \right)
  + \Omega_h,
\end{equation}
where $\Omega_h$ is the contribution of the hadrons, $\Omega_i$ that
of charged leptons and $\Omega_{\nu_i}$ that of the neutrinos. To a
good approximation the leptons can be described by a free
non-interacting gas of fermions. Hence the thermodynamic potential of
the charged leptons is given by
\begin{equation}
 \Omega_i =  - \frac{T}{\pi^2}
  \sum_{\pm} \int_0^\infty \mathrm{d}p \, p^2
  \log \left[ 1 + e^{-\beta (\omega_p \pm \mu_i)} \right],
\end{equation}
where $\omega_p = (p^2 + m_i^2)^{1/2}$ and $\beta = 1/T$ denotes
the inverse temperature.  For the neutrinos (which only have
left-handed chirality) one has
\begin{equation}
 \Omega_{\nu_i} =  - \frac{T}{2\pi^2}
 \sum_{\pm}  \int_0^\infty \mathrm{d}p \, p^2
  \log \left[ 1 + e^{-\beta (p \pm \mu_{\nu_i})} \right].
\end{equation}

The densities of the leptons can be found by taking the derivative of
the thermodynamical potential with respect to the chemical potential.
In this way for zero temperature one finds that the density of the
charged leptons equals
\begin{equation}
 n_i =
\frac{1}{3\pi^2}
\mathrm{sgn}(\mu_{L_i} - \mu_Q)
 \left [
 (\mu_{L_i} - \mu_Q)^2 - m_i^2 \right]^{3/2},
\label{eq:fermiondensity}
\end{equation}
for $\vert \mu_{L_i} - \mu_Q \vert \geq m_i$ and $n_i = 0$
otherwise. The density of the neutrinos is equal to
\begin{eqnarray}
 n_{\nu_i} =  \frac{1}{6\pi^2} \mu_{L_i}^3 + \frac{1}{6} \mu_{L_i} T^2.
\end{eqnarray}

Now let us set the temperature to zero and take $\mu_B$ small so that
there are no baryons in the system. This means that the chemical
potential of the proton ($\mu_p = \mu_B + \mu_Q$) should be smaller
than the proton mass $m_p$.  Since $\mu_Q$ can become $m_\pi$, this
translates into the requirement that $\vert \mu_B \vert \lesssim m_p -
m_\pi$. In that case it is well known from the analysis using the
chiral effective Lagrangian that a $\pi^+$ condensate will form via a
second-order transition if $\mu_Q > m_\pi$ \cite{Son01, Kogut01}. This
is confirmed by lattice QCD simulations \cite{Kogut02}.

If in this particular situation $\vert \mu_Q \vert $ is smaller than
$m_\pi$, there are no hadronic states around. Hence the hadronic
sector is automatically electrically neutral. The question then is,
can we neutralize the leptonic sector such that $\mu_Q = m_\pi$?  If
so, we have found the onset of pion condensation in electrically
neutral matter.  Answering this question amounts to solving the
following equation
\begin{equation}
\sum_i n_i(\mu_Q = m_\pi) = 0. \label{eq:onsetpion}
\end{equation}
If there are no neutrinos around this equation has no solution since
$\mu_Q = m_\pi$ automatically introduces a nonzero density of
electrons and muons.  However, if we for example take $\mu_{L_e} =
\mu_{L_\mu} \equiv \mu_L$ we find that $\mu_Q = \mu_L$ guarantees
electric neutrality. Pion condensation then sets in at $\mu_L =
m_\pi$. This corresponds to a electron and muon neutrino number
density of $m_{\pi}^3 / (6 \pi^2) \approx 5.9 \times
10^{-3}\;\mathrm{fm}^{-3}$.

\subsection{Phase diagram $\mu_{L_e}$ vs.\  $\mu_{L_\mu}$ at $T=0$}
\begin{figure}[t]
\includegraphics[scale=1]{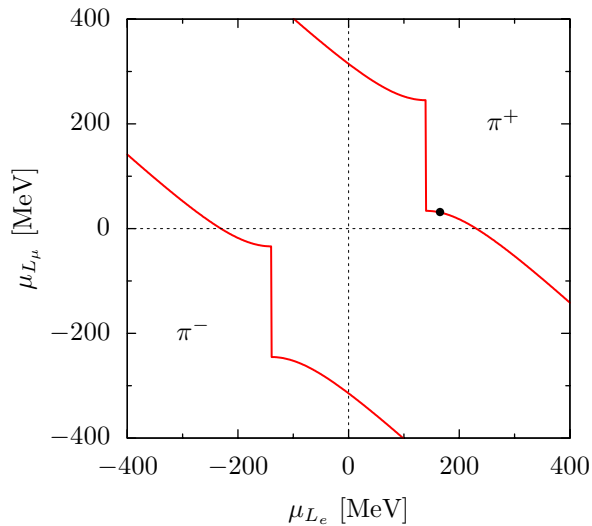}
\caption{Phase diagram of electrically neutral QCD at $T=0$,
valid for $\vert \mu_B \vert \lesssim m_p -  m_\pi$,
 as a function of the electron and muon lepton number chemical potential.
The phase boundary denotes a second-order transition. The black dot
denotes the point where $n_{L_\mu} = 0$.
\label{fig:mulemulmu}}
\end{figure}

Now that we found the onset of pion condensation in electrically
neutral matter for equal electron and muon lepton number chemical
potential, let us see what happens in the more general case in which
they are different. By solving Eq.~(\ref{eq:onsetpion}) for $\mu_Q =
m_\pi$ we obtain the phase diagram of QCD as a function of $\mu_{L_e}$
and $\mu_{L_\mu}$ for $T=0$ and small baryon chemical potential. We
display the result in Fig.~\ref{fig:mulemulmu}.

Let us discuss the transition line to the $\pi^+$-condensed phase in
somewhat more detail. If $m_\pi - m_\mu < \mu_{L_\mu} < m_\pi + m_\mu$
there are no muons in the system so it should be neutralized solely by
the electrons. This then gives $m_\pi - m_e < \mu_{L_e} < m_\pi + m_e$
which corresponds to the vertical lines in the phase diagram of
Fig.~\ref{fig:mulemulmu}. Hence the length of the vertical line is
equal to twice the muon mass. For $\mu_{L_e} < m_\pi - m_e$ the exact
solution to the neutrality constraint gives the following phase
boundary
\begin{equation}
 \mu_{L_\mu} = m_\pi + \sqrt{(\mu_{L_e} - m_\pi)^2 - m_e^2 + m_\mu^2}.
\end{equation}
while for  $\mu_{L_e} > m_\pi + m_e$ we find
\begin{equation}
 \mu_{L_\mu} = m_\pi - \sqrt{(\mu_{L_e} - m_\pi)^2 - m_e^2 + m_\mu^2}.
\end{equation}
If the muon lepton number chemical potential vanishes, a $\pi^+$
condensate will form if
\begin{equation}
\mu_{L_e} > m_\pi + \sqrt{m_\pi^2 + m_e^2 - m_\mu^2} = 231\;\mathrm{MeV}.
\end{equation}
This corresponds to an electron neutrino density of
 $\mu_{L_e}^3/(6 \pi^2) = 2.7 \times 10^{-2}\; \mathrm{fm}^{-3}$.

In some situations like the proto-neutron star it is natural to
require zero muon lepton number density ($n_{L_\mu} = n_\mu +
n_{\nu_\mu} = 0$).  In such a case we find that the transition to the
pion condensed phase occurs at
\begin{equation}
 \mu_{L_\mu} = \xi m_\pi - \xi \sqrt{m_\pi^2 + (m_\mu^2 - m_\pi^2) / \xi}
= 31\;\mathrm{MeV},
\label{eq:onsetMU}
\end{equation}
here $\xi = 1 / [1- (1/2)^{2/3}]$.
The electron chemical potential is then equal to
\begin{equation}
 \mu_{L_e} = m_\pi + \sqrt{ (\tfrac{1}{2})^{2/3} \mu_{L_\mu}^2 + m_e^2}.
\end{equation}
To a good approximation we can neglect the electron mass in the
last equation. Then by expanding Eq.~(\ref{eq:onsetMU}) in
powers of $(m_\pi^2 - m_\mu^2)/m_\pi^2$ we obtain
\begin{equation}
 \mu_{L_e} \approx m_\pi \left( 1 + \frac{1}{2^{4/3}}
  \frac{m_\pi^2 - m_\mu^2}{m_\pi^2}
\right) = 164\;\mathrm{MeV}.
\end{equation}
The solution using the exact expression gives $\mu_{L_e} = 163\;
\mathrm{MeV}$.  Hence if the muon lepton number density vanishes pion
condensation will occur if the electron neutrino density is larger
than $\mu_{L_e}^3/(6 \pi^2) = 9.6 \times 10^{-3}\; \mathrm{fm}^{-3}$.

\subsection{Phase diagram $\mu_{L_e}$ vs.\  $\mu_{L_\mu}$ for $T \ll m_\pi$}
Let us now discuss how the $\mu_{L_e}$ vs.\ $\mu_{L_\mu}$ phase
diagram of electrically neutral QCD looks like at finite temperature.
As long as the temperature is much smaller than the chiral phase
transition temperature $T_c$, the mass of the pion is not much
different from the vacuum pion mass $m_\pi$.  Hence for $T \ll T_c$
the onset of pion condensation still occurs to a good approximation at
$\mu_Q = m_\pi$.

\begin{figure}[t]
\includegraphics[scale=1]{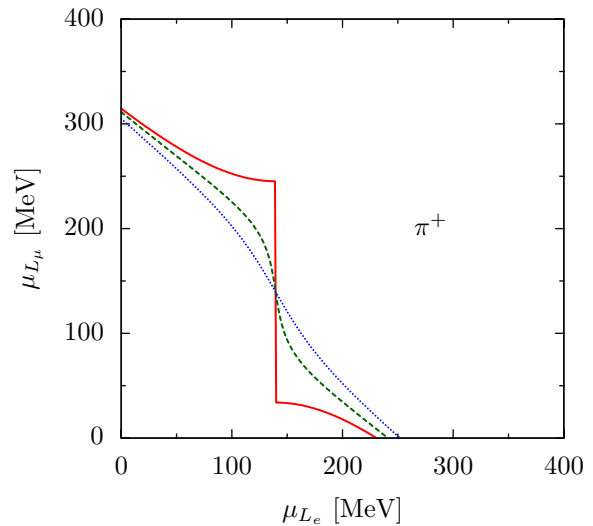}
\caption{Phase diagram of electrically neutral QCD, valid for $\vert
 \mu_B \vert \lesssim m_p - m_\pi$, for $T = 0$ (solid, red), $25$
(dashed, green), and $50\;
\mathrm{MeV}$ (dotted, blue), as a function of the electron and muon lepton
  number chemical potential. The $T=25$ and $50\;\mathrm{MeV}$ results are
estimates.}
\label{fig:mulemulmuT}
\end{figure}

If the temperature is also much smaller than $m_\pi$, there are to a
good approximation no hadronic states in the system as long as $\mu_Q
< m_\pi$. Hence we can like in the previous subsection find the onset
of pion condensation by solving Eq.~(\ref{eq:onsetpion}), but now at
finite temperature. We display the results in
Fig.~\ref{fig:mulemulmuT}.

To understand Fig.~\ref{fig:mulemulmuT} better, let us see what
happens if the temperature is increased. In that case the muon mass,
which is equal to half of the length of the vertical line in the phase
diagram, becomes less important. Let us for a moment neglect the muon
mass completely.  Then the solution of Eq.~(\ref{eq:onsetpion}) is a
straight line from $(\mu_{L_e}, \mu_{L_\mu}) = (2m_\pi, 0)$ to $(0,
2m_\pi)$. At these high temperatures where the muon mass can be
neglected, the approximation of ignoring the hadrons is wrong, so that
the straight line will never be the real phase boundary.
Nevertheless, this argument shows that by raising the temperature the
tendency of the phase boundary is to approach the straight line as
long as $T \ll m_\pi$ in agreement with the results displayed in
Fig.~\ref{fig:mulemulmuT}.  As a result of this straightening tendency
there are points in the phase diagram ($\mu_{L_e} < m_\pi$,
$\mu_{L_\mu} > m_\pi$) where the pion condensate does not appear
at $T=0$ but arises if the temperature is increased.

\section{NJL model calculation}
As was shown in the previous section, pion condensation arises in
electrically neutral hadronic matter for large neutrino densities.  We
have obtained the phase diagram of QCD for low temperatures and
small baryon chemical potential. It is of interest to see what happens
to the phase diagram at larger baryon chemical potential and higher
temperatures.

To investigate the phase diagram at finite $T$ and $\mu_B$ we need to
know the hadronic component of the thermodynamic potential, $\Omega_h$.
Unfortunately we can not compute $\Omega_h$ in QCD from first
principles. We will resort to an effective model of QCD, the
Nambu--Jona-Lasinio (NJL) model (see e.g. Ref.~\cite{Buballa_RPT} for
an extensive review). The NJL model qualitatively describes QCD.  It
for example captures the chiral phase transition and can be used to
study mesons and pion condensation. However, it is important to
realize that nuclear matter is not described at all with this model.
The phase diagrams obtained with the NJL model will therefore only
describe the qualitative features of the QCD phase diagram.
Nevertheless, as we argued in the previous section, our conclusion
that a large part of the phase diagram contains the pion condensate is
certainly also correct for QCD.

The 2-flavor NJL model we will use in this article is given
by the following Lagrangian density
\begin{equation}
\begin{split}
\mathcal{L} & =\bar \psi \left(i \gamma^\mu \partial_\mu - m
  + \mu \gamma_0 \right)\psi
+(1-\alpha)\mathcal{L}_1+\alpha\mathcal{L}_2,
\\
\mathcal{L}_1 & =G[(\bar\psi\psi)^2+(\bar\psi\vec\tau\psi)^2
+(\bar\psi i\gamma_5\psi)^2+(\bar\psi i\gamma_5\vec\tau\psi)^2],
\\
\mathcal{L}_2 & =G[(\bar\psi\psi)^2-(\bar\psi\vec\tau\psi)^2
-(\bar\psi i\gamma_5\psi)^2+(\bar\psi i\gamma_5\vec\tau\psi)^2].
\end{split}
\label{eq:lagrnjl}
\end{equation}
Here $G$ is the coupling constant and $\vec\tau$ denotes the Pauli
matrices. The mass matrix $m$ is diagonal and contains the bare quark
masses $m_{u}$ and $m_{d}$. The matrix $\mu$ is also diagonal and
contains the quark chemical potentials $\mu_u$ and $\mu_d$, which are
given in Eq.~(\ref{eq:chempot}). The interaction terms $\mathcal L_1$
and $\mathcal L_2$ are invariant under $\mathrm{SU(2)_L\times
  SU(2)_R}$ transformations, as is QCD.  The interaction $\mathcal
L_1$ preserves the axial $\mathrm{U(1)_A}$ symmetry, while $\mathcal L_2$
(the two-flavor 't~Hooft term) breaks it. In QCD the $\mathrm{U(1)_A}$
symmetry is broken, so $\alpha$ should be nonzero.

Since the NJL model is an effective model which is only valid for low
momenta, one has to specify an ultraviolet regularization. In this
paper we will employ a three-dimensional sharp momentum cutoff
$\Lambda$.

We will take the following parameter set which is also used in other
papers (see e.g.~Refs.~\cite{Ratti06, Zhang08}): $G =
5.04\;\mathrm{GeV}^{-2}$, $\Lambda = 651\;\mathrm{MeV}$ and $m \equiv
m_{u} = m_{d} = 5.5\;\mathrm{MeV}$.  We will treat $\alpha$ as a free
parameter.  At zero temperature and baryon chemical potential, its
most likely value lies somewhere between $0.1$ and $0.2$
\cite{Buballa_RPT, Frank2003}.  With this parameter set one can
reproduce the vacuum properties of the mesons, $m_{\pi^\pm} \approx
140\;\mathrm{MeV}$, $f_\pi \approx 92\;\mathrm{MeV}$.

In order to compute the thermodynamic potential, we will resort to the
mean-field approximation. We will introduce the following mean fields
\begin{equation}
\begin{split}
 \sigma_u &= - 4 G \langle \bar u u \rangle, \\
 \sigma_d &= - 4 G \langle \bar d d \rangle, \\
 \rho &= - 2 G \langle \bar \psi \tau_2 i \gamma_5 \psi \rangle,
\end{split}
\label{eq:condensates}
\end{equation}
here $\sigma_u$ and $\sigma_d$ are proportional to the chiral
condensates, while the pion condensate is described by $\rho$.

The hadronic part of the thermodynamic potential can now be found by
expanding the Lagrangian density around the mean fields and
integrating out the fermions. In that way one obtains
\begin{multline}
 \Omega_h =
\frac{\tfrac{1}{2} (1-\alpha) (\sigma_u^2 + \sigma_d^2) + \alpha
 \sigma_u \sigma_d + \rho^2}{4 G}
\\
-
 \frac{N_c}{2\pi^2}  \sum_{i=1}^{4} \int_0^\Lambda
 \mathrm{d} p\, p^2
\left [
\vert \lambda_i \vert
+
2 T \log \left(
1 + e^{-\beta \vert \lambda_i \vert}
\right)
\right]
.
\end{multline}
Here $N_c = 3$ denotes the number of colors and $\lambda_i$ are the
four independent eigenvalues of the mean-field Hamiltonian
$\mathcal{H}$ which is given by
\begin{equation}
 \mathcal{H} =
\left (
\begin{array}{cc}
\gamma^0 \vec \gamma \cdot \vec p + M_u \gamma_0 - \mu_u
& \gamma^0 \gamma_5 \rho \\
- \gamma^0 \gamma_5 \rho &
\gamma^0 \vec \gamma \cdot \vec p + M_d \gamma_0 - \mu_d
\end{array}
\right),
\end{equation}
where the constituent quark masses are equal to
\begin{equation}
\begin{split}
M_u &= m_{u} +
(1 - \alpha) \sigma_u + \alpha \sigma_d,\\
M_d &= m_d + (1-\alpha) \sigma_d + \alpha \sigma_u.
\end{split}
\end{equation}

\begin{figure*}[t]
\includegraphics[scale=1]{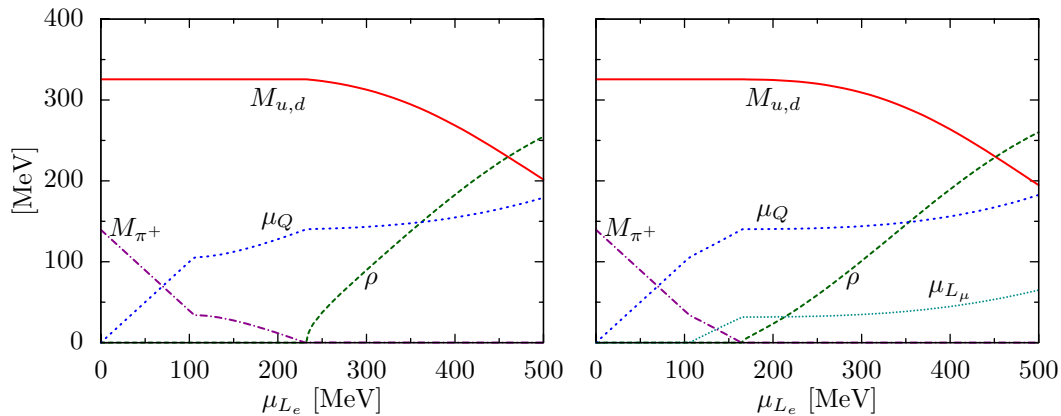}
\caption{Evolution of the condensates and chemical potentials along
  the $\mu_B=0$ axis at zero temperature. Left panel: The
  $\mu_{L_\mu}=0$ case. Right panel: The $n_{L_\mu}=0$ case.
  Displayed are: constituent quark mass (red, solid), pion condensate
  (green, long-dashed), electric charge chemical potential (blue,
  short-dashed), in-medium $\pi^+$ mass (magenta, dash-dotted), and muon lepton
  number chemical potential (cyan, dotted). The in-medium pion mass
  is discussed in Sec.~\ref{sec:pionmass}.}
\label{Fig:chirotation}
\end{figure*}

If $M_u = M_d \equiv M$ the four eigenvalues can be obtained
analytically and read
\begin{equation}
\lambda_i  =
\sqrt{ (\omega_p \pm \mu_Q / 2)^2 + \rho^2} \pm \bar \mu
\label{gapped_spectrum}
\end{equation}
where $\bar \mu = \mu_B / 3 + \mu_Q / 6$ and $\omega_p = \sqrt{p^2 +
  M^2}$.  The values of the mean fields follow by minimizing the
thermodynamic potential with respect to these mean fields, which
amounts to solving the following equations
\begin{equation}
 \frac{\partial \Omega}{\partial \sigma_u} =
 \frac{\partial \Omega}{\partial \sigma_d} =
 \frac{\partial \Omega}{\partial \rho} = 0.
 \label{eq:gapeqs}
\end{equation}
In addition one has to solve the electric neutrality constraint,
Eq.~(\ref{eq:neutrality}), and in the case where we require zero muon
lepton number density, also the constraint $n_{L_\mu} = 0$. If $\alpha
= 1/2$, the effective potential is a function of $\sigma_u +
\sigma_d$. Hence in that case one always finds $\sigma_u = \sigma_d$
as a solution. For other values of $\alpha$, $\sigma_u$ and $\sigma_d$
can be different.

We have solved Eq.~(\ref{eq:gapeqs}) and the neutrality constraints
numerically and will discuss the results in the following subsections.

\subsection{Condensates at $T=0$ and $\mu_B=0$}
In Fig.~\ref{Fig:chirotation} we show the results achieved at zero
temperature and baryon chemical potential. In this limit the
condensation is governed by the chiral physics and the NJL model,
augmented with the gas of free leptons to account for charge
neutrality, is therefore expected to be most reliable.

First of all, note that the NJL calculation naturally reproduces the
previous model-independent evaluation of the onset of pion
condensation. To demonstrate that even the quantitative results inside
the pion-condensed phase are to a large extent model-independent, one
can employ chiral perturbation theory \cite{Son01, Kogut01}. The
leading-order (Euclidean) chiral Lagrangian reads
\begin{equation}
\mathcal L_{\mathrm{\chi PT}}=\frac{f_\pi^2}{4}\left[
\mathrm{Tr}(D_\mu U^\dag D_\mu U)- 2 m_\pi^2\mathrm{Re\,Tr}U
\right],
 \label{eq:chpt}
\end{equation}
where $U$ is a unitary $2\times2$ matrix field and the covariant
derivative, $D_\mu U=\partial_\mu U-\ell_\mu U+Ur_\mu$, incorporates
its coupling to external left- and right-handed vector fields,
$\ell_\mu$ and $r_\mu$. In presence of electric charge or isospin
chemical potential, they are $\ell_0=r_0=\frac12\tau_3\mu_Q$.

The ground-state expectation value of the order parameter can be without
lack of generality cast as
\begin{equation}
U=\begin{pmatrix}
\cos\theta & \sin\theta\\
-\sin\theta & \cos\theta
\end{pmatrix}.
\end{equation}
If $\theta=0$ we recover the vacuum chiral condensate, while
$\theta=\pi/2$ would be a purely pion condensate. In general this
rotation angle is determined by minimization of the static part of the
Lagrangian and for $|\mu_Q|>m_\pi$ is given by \cite{Son01}
\begin{equation}
\cos\theta=\frac{m_\pi^2}{\mu_Q^2}.
\end{equation}
This is analogous to the gap equation in the NJL model and provides a
model-independent relation among the chiral and pion condensates and
the electric charge chemical potential. The electric charge density of
the pion condensate is in turn obtained as \cite{Son01}
\begin{equation}
n_\pi= -\frac{\partial\mathcal L_{\mathrm{\chi PT}}}{\partial \mu_Q}
 = f_\pi^2\mu_Q\biggl(1-\frac{m_\pi^4}{\mu_Q^4}\biggr).
\end{equation}

We used these expressions supplemented with the lepton sector to check
the results of Fig.~\ref{Fig:chirotation}. Since chiral perturbation
theory allows a straightforward evaluation of the expectation values
of fermion bilinear operators rather than the NJL order parameters
$\sigma_{u,d},\rho$, see Eq.~(\ref{eq:condensates}), we use the vacuum
NJL value of $\sigma_u=\sigma_d$ to normalize the condensates obtained
with chiral perturbation theory. The results agree with NJL model up
to about a $2\%$ accuracy even at the highest values of $\mu_{L_e}$
considered. Therefore, we do not plot the corresponding lines in
Fig.~\ref{Fig:chirotation}; they would hardly be distinguishable.
(See \cite{He05b} for a similar comparison.)

\subsection{Phase diagram $\mu_{L_e}$ vs. $\mu_{L_\mu}$
 at finite temperature}
\begin{figure}[t]
\includegraphics{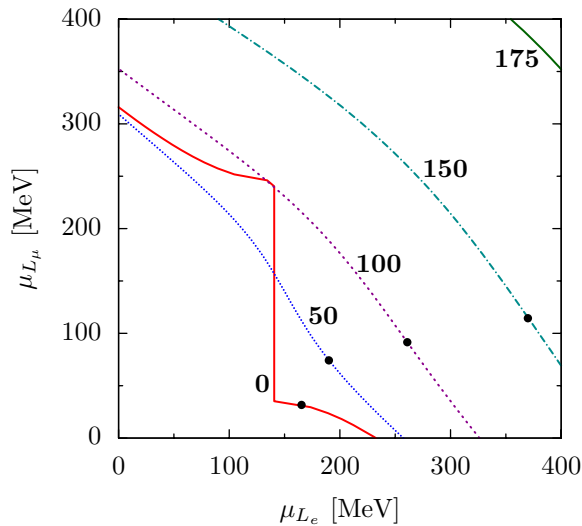}
\caption{Phase diagram of the electrically neutral NJL model  with
    $\alpha = 0$ at various values of temperature (indicated by the
    numbers in boldface).  The black dots denote the points at the
  different phase boundaries where the muon lepton number density
  ($n_{L_\mu}$) vanishes.
  \label{fig:NJLmulemulmuT}}
\end{figure}

In Fig.~\ref{fig:NJLmulemulmuT} we display the phase diagram of the
NJL model as a function of $\mu_{L_e}$ and $\mu_{L_\mu}$ for different
temperatures at $\mu_B = 0$ and $\alpha = 0$.  We found that up to
about $T=150\;\mathrm{MeV}$ this phase diagram does not change much by
modifying $\alpha$.  Comparing this phase diagram to
Fig.~\ref{fig:mulemulmuT} we find very good agreement with the
model-independent estimates up to $T=50\;\mathrm{MeV}$. Hence we can
say that Fig.~\ref{fig:mulemulmuT} most likely represents the phase
diagram of electrically neutral QCD for small baryon density as a
function of lepton number chemical potentials.

For higher temperatures the NJL model calculation serves as an
illustration of how the phase diagram of QCD would qualitatively look
like. We find that the onset of the pion condensed phase moves to
higher values of the lepton number chemical potentials. This is caused
by the appearance of charged hadronic states in the system which
affects the neutrality.  Furthermore due to the increase of the pion
mass at finite temperature, the onset of pion condensation moves to
larger values of $\mu_Q$.

\subsection{Phase diagram $\mu_B$ vs.\ $\mu_{L}$ at zero temperature}
\begin{figure*}[t]
\includegraphics[scale=1]{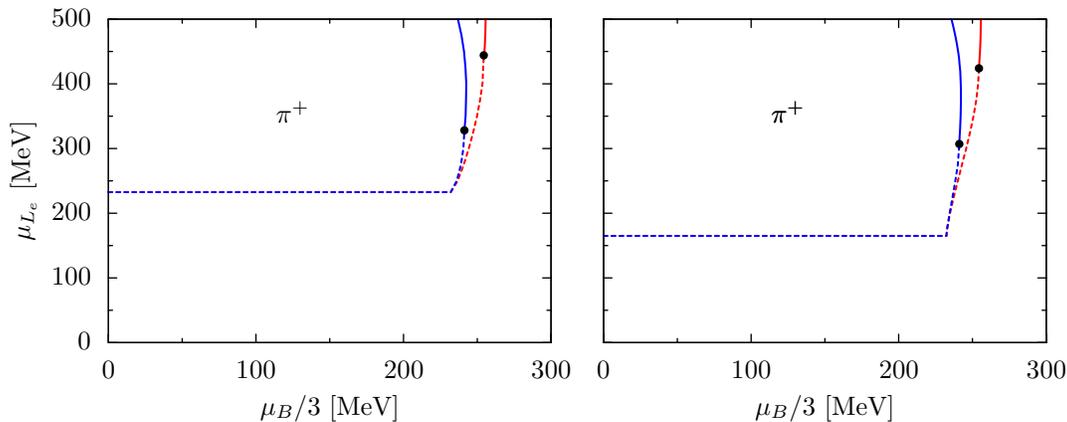}
\caption{Phase diagram of the electrically neutral NJL model in the
  $\mu_B-\mu_{L_e}$ plane at $T=0$. Left panel: The $\mu_{L_\mu}=0$ case. Right
  panel: The $n_{L_\mu}=0$ case. The solid and dashed lines denote
  first and second order phase transitions, respectively. In each plot
  we display two curves corresponding to $\alpha=0$ (left, blue) and
  $\alpha=1/2$ (right, red). The black dots correspond to critical
  points.}
\label{Fig:PDmuLmuB}
\end{figure*}

In Ref.~\cite{Ruester06} the phase diagram of neutral matter with
neutrinos present was discussed for very large baryon chemical
potential where color superconducting phases arise. See also
Refs.~\cite{Laporta06, Sandin07}.  Here we complement this phase
diagram with the results obtained at low baryon chemical potential. We
display the results in Fig.~\ref{Fig:PDmuLmuB} for the two cases
regarding the muon-lepton content discussed above, and for two
different values of the $\alpha$ parameter.

The structure of the phase diagram is simple. The straight line, which
determines the onset of pion condensation at low baryon chemical
potential, is given by the model independent argument presented in
Sec.~II. The corner of the pion condensed phase, where this line
breaks, is marked by the appearance of up quarks in the system. In the
NJL model at the mean-field approximation, quarks behave as
noninteracting quasiparticles with effective energy gap (mass)
determined by $M_{u,\text{eff}}=M_u-\frac13\mu_B-\frac23\mu_Q$,
$M_{d,\text{eff}}=M_d-\frac13\mu_B+\frac13\mu_Q$. When this drops
below zero, a Fermi sea of quarks is formed. This happens along the
boundary of the pion condensed phase, when
$\frac{\mu_B}3=M_u-\frac23m_\pi$. Using the constituent quark mass in
the vacuum, $M_{u,d} = 325\;\mathrm{MeV}$ for our parameter set,
we find $\frac{\mu_B}3 = 232\;\mathrm{MeV}$.

\begin{figure}[t]
\includegraphics[scale=1]{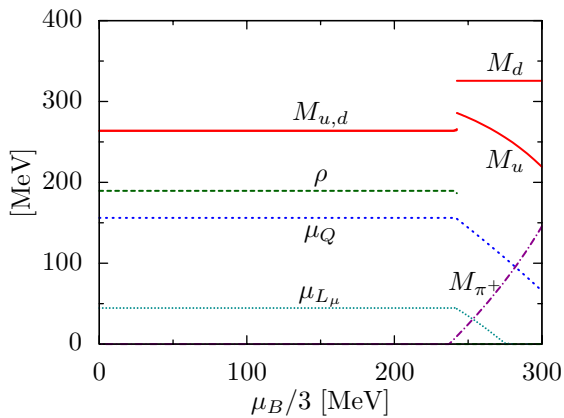}
\caption{Evolution of the condensates, chemical potentials and the
  in-medium pion mass as a function of $\mu_B$ for
  $\mu_{L_e}=400\;\mathrm{MeV}$, $\alpha=0$ and $T=0$ in the $n_{L_{\mu}}=0$
  case.}
\label{Fig:section400}
\end{figure}

The line of second order phase transition eventually ends up in a
critical point from which on the transition becomes first order. To
demonstrate this, we plot in Fig. \ref{Fig:section400} the values of
the condensates, the charge chemical potentials, and the in-medium
pion mass (obtained as the pole of the pion propagator \cite{Abuki08})
along the section of the phase diagram with
$\mu_{L_e}=400\;\mathrm{MeV}$ for the $n_{L_{\mu}}=0$ case and
$\alpha=0$. We observe that as $\alpha$ decreases, the critical point
moves down and the pion-condensed phase shrinks to somewhat smaller
$\mu_B$. This is because small $\alpha$ tends to split the constituent
masses of up and down quarks and thus disfavor the pairing mechanism
which underlies the pion condensation.

\section{Pion properties in a neutrino gas}
As we saw in the previous sections, the lepton medium induces an
isospin chemical potential. This chemical potential and the finite
density of neutrinos will modify the behavior of the pions. Let us
therefore have a closer look at the spectral properties of the pions
in the lepton medium, in particular their masses and decay rates.

\subsection{Masses}\label{sec:pionmass}
As long as the isospin is not spontaneously broken (i.e., there is no
pion condensate), the in-medium pion masses are simply given by
$M_{\pi^\pm}=m_\pi\mp\mu_Q$ \cite{Son01, Kogut01}. The pion
condensation sets in where one of the masses drops to zero.

At low $\mu_{L_e}$, only electrons are light enough to be excited. In
order to preserve electric neutrality, we have to make sure that the
electron chemical potential is zero, that is, $\mu_Q=\mu_{L_e}$. The
pion masses are thus simply equal to $M_{\pi^\pm}=m_\pi\mp\mu_{L_e}$.
When $\mu_{L_e}$ exceeds the muon mass, neutrality becomes a
nontrivial issue and the pion masses depend on the exact way it is
imposed. We will describe in detail two special cases: Zero muon lepton
number chemical potential, and zero muon lepton number density, both
of which were already discussed in Sec. II.

At $\mu_{L_\mu}=0$ we have $\mu_{\mu^\pm}=\pm\mu_Q$ so that for
$\mu_{L_e}>m_\mu+m_e$, antimuons will appear in the system. In this
case electric neutrality requires that the electrons and antimuons
have the same Fermi momentum, which leads to
\begin{equation}
\mu_Q=\frac{\mu_{L_e}^2+m_\mu^2-m_e^2}{2\mu_{L_e}}.
\end{equation}
Setting $\mu_Q=m_\pi$ recovers the transition point to the pion condensed
phase, Eq. (11).

If we instead demand zero muon lepton number density, the $\mu^+$
chemical potential modifies to $\mu_Q-\mu_{L_\mu}$ and $\mu_{L_\mu}$
is determined self-consistently from the muon lepton number neutrality
condition; a negative contribution from antimuons has to be
compensated by a finite density of muon neutrinos. Solving the set of
two neutrality equations, we obtain the results shown in Fig.
\ref{Fig:chirotation} by the dash-dotted lines. The mass of $\pi^-$ is
given simply by a reflection of the curves with respect to the $M=m_\pi$ line.

\begin{figure}
\includegraphics[scale=1]{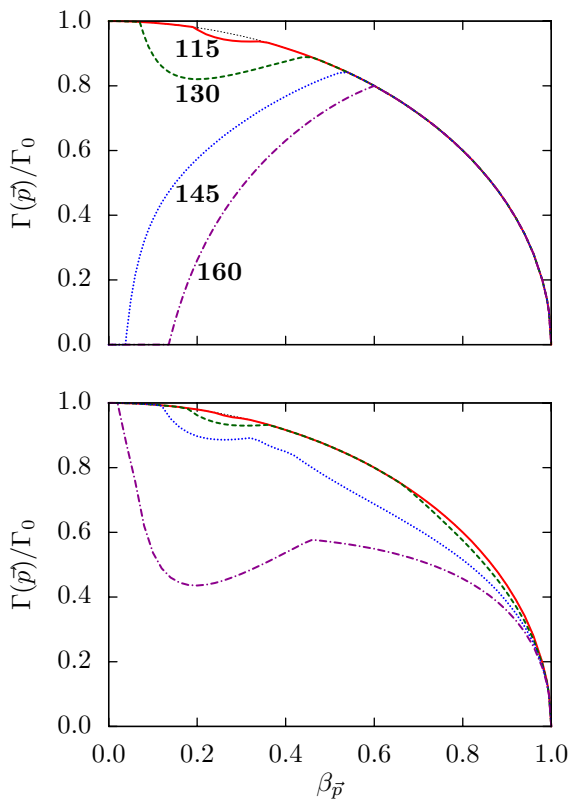}
\caption{Decay rate of $\pi^+$ in units of the vacuum rate at rest as
  a function of the pion velocity for various values of electron
  lepton number chemical potential (indicated by the numbers in
  boldface; the same values are implied in the lower plot). The thin
  dotted line indicates the vacuum decay rate, i.e., is trivially
  given by the $\gamma$-factor. Upper panel: The $\mu_{L_\mu}=0$ case.
  Lower panel: The $n_{L_\mu}=0$ case.}
\label{Fig:decay}
\end{figure}

\subsection{Decay rates}
The charged pions in the vacuum decay predominantly in the
$\pi^+\to\mu^+\nu_\mu$, $\pi^-\to\mu^-\bar\nu_\mu$ channels. These
constitute about $99.99\%$ of the total decay rate \cite{PDG}. In the
vacuum and at rest, the decay rate is, at the leading order in weak
interactions, given by the textbook formula
\begin{equation}
\Gamma_0=\frac1{4\pi}V_{ud}^2G_{\text{F}}^2f_\pi^2m_\pi m_\mu^2\biggl(1-
\frac{m_\mu^2}{m_\pi^2}\biggr)^2.
\end{equation}
Here $V$ denotes the Cabibbo--Kobayashi--Maskawa matrix and
$G_{\mathrm{F}}$ the Fermi constant.  The presence of the medium
breaks explicitly Lorentz invariance, and we will therefore calculate
the decay rates as a function of momentum.  The derivation entails two
ingredients which can be considered separately: The invariant decay
amplitude and the kinematics.

To determine the invariant amplitude, we need to know the coupling of
the pion to the charged weak current. In vacuum, this is equal to
$\frac i2gV_{ud}f_\pi p^\mu$, where $p^\mu$ is the pion four-momentum
and $g$ the weak coupling constant. To determine how this is modified
in the medium, we use chiral perturbation theory \cite{Scherer}.
Introducing in the lowest order pion effective Lagrangian
(\ref{eq:chpt}) the external vector electromagnetic as well as the
left-handed charged weak current, we find that one simply has to make
the replacement $p^\mu\to\tilde p^\mu=(p_0+\mu_Q,\vec p)$. In this
simple estimate of the decay rate, we neglect truly quantum
corrections to the pion decay constant which give it a weak medium
dependence \cite{Barducci}.

An explicit calculation shows that this change of the pion coupling to
the weak current is precisely what is needed to make the total
amplitude for the pion decay completely independent of the chemical
potentials; the extra term in the pion--weak-current coupling cancels
with a similar term coming from the muon and neutrino
wave functions. As a consequence the kinematics becomes trivial: The
integration over phase space for the final state can be transformed
into the center-of-mass frame (CMS). The only effect of the medium is
a restriction of the phase space due to the fact that some of the
momentum states are occupied by particles in the Fermi sea. The final
formula for the decay rate as a function of momentum $\vec p$ then
reads
\begin{equation}
\frac{\Gamma(\vec p)}{\Gamma_0}=\frac1{\gamma_{\vec
    p}}\frac{\omega^*}{4\pi},
\end{equation}
where $\gamma_{\vec p}$ denotes the Lorentz factor and $\omega^*$ is
the allowed solid angle for the products of the decay, measured in
CMS.

First of all, observe the decay rate of $\pi^-$ is trivially the same
as in vacuum, because there are neither muons nor muon antineutrinos
in the system.  In the following we will therefore concentrate on the
decay of $\pi^+$. In CMS the magnitude of the momentum of the antimuon
and muon neutrino from the $\pi^+$ decay is
\begin{equation}
k_0=\frac{m_\pi^2-m_\mu^2}{2m_\pi};
\end{equation}
the decay is isotropic. (Note that since both the electric charge and
the lepton number are preserved in the decay, the presence of chemical
potentials does not affect the energy and momentum conservation.)
Therefore, when the Fermi momentum of both antimuons and neutrinos is
smaller than $k_0$, the decay rate will be $\Gamma_0$, otherwise it
will be zero, at least for the considered process.

The energies of the antimuon and the neutrino in the rest frame of the medium
are given by a simple Lorentz boost,
\begin{equation}
\epsilon_\mu=\gamma_{\vec p}(\epsilon_0+\beta_{\vec
  p}k_0\cos\theta^*),\quad \epsilon_\nu=\gamma_{\vec
  p}k_0(1-\beta_{\vec p}\cos\theta^*),
\end{equation}
where $\beta_{\vec p}$ denotes the velocity. Furthermore,
$\epsilon_0=(m_\pi^2+m_\mu^2)/(2m_\pi)$ is the antimuon energy and
$\theta^*$ the angle of the antimuon with respect to the pion
momentum, both measured in CMS. This allows a straightforward
determination of the available phase space for the decay. For
instance, in the case $\mu_{L_\mu}=0$, only phase space blocking by
antimuons occurs (there are no muon neutrinos) and one may explicitly
express $\omega^*$ as
\begin{equation}
\frac{\omega^*}{4\pi}=\frac12\left[1-\Xi\left(\frac1{\beta_{\vec
p}k_0}\Bigl(\frac{\mu_Q}{\gamma_{\vec p}}-\epsilon_0\Bigr),-1,+1\right)\right],
\end{equation}
where the value of the function $\Xi(x,a,b)$ with $a<b$ is equal to
$a$ if $x<a$, to $b$ if $x>b$, and to $x$ otherwise, that is,
$\Xi(x,a,b)=\min[\max(x,a),b]$.

The numerical results for the decay rate as a function of the pion
velocity are shown in Fig.~\ref{Fig:decay}. The interpretation of the
results is simple. If the chemical potentials are low enough such that
the decay is not blocked in CMS, then boosting to finite momentum may
bring the backward emitted particles into the Fermi sea and thus
suppress the decay rate beyond the simple $\gamma$-factor from time
dilation. On the other hand, if the decay is Pauli-blocked in CMS,
boosting to finite momentum may liberate the forward-emitted particles
so that the decay becomes possible. In the $n_{L_\mu} = 0$ case parts
of the phase space are blocked by both antimuons and muon neutrinos.

\section{Conclusions}
In this article we have shown that a positively charged pion
condensate arises in electrically neutral matter at high neutrino and
small baryon densities. We found that at zero temperature, zero baryon
chemical potential, and zero muon lepton number density, the onset of
pion condensation lies at an electron neutrino number density of $9.6
\times 10^{-3}\; \mathrm{fm}^{-3}$. 

For zero temperature we have obtained the phase diagram of
electrically neutral QCD as a function of $\mu_{L_e}$ and
$\mu_{L_\mu}$, valid for small baryon chemical potential, $\vert
\mu_B \vert \lesssim m_p - m_\pi$. We have estimated this phase
diagram at higher temperatures. Comparison to model calculations with
obtained using the NJL model show that up to about
$T=50\;\mathrm{MeV}$ these estimates are very good.

Using the NJL model we also studied the behavior of the phase diagram
as a function of baryon chemical potential and lepton number chemical
potential. We found that the pion condensed phase arises up to $\vert
\mu_B/3 \vert \approx M - \frac23 m_\pi$, where $M$ is the
constituent quark mass in the vacuum.

Hence the main conclusion of this work is that the pion condensed
phase makes up a large part of the phase diagram of electrically
neutral QCD at finite lepton number chemical potential and small
baryon chemical potential.

\section*{Acknowledgments}
We would like to thank J.O.~Andersen, D.~Boer and D.H.~Rischke for
discussions.  The work of H.A.\ and T.B.\ is supported by the
Alexander von Humboldt Foundation. H.J.W. acknowledges the support of
the Helmholtz Alliance Program of the Helmholtz Association, contract
HA-216 `Extremes of Density and Temperature: Cosmic Matter in the
Laboratory'.

\appendix

\end{document}